\documentclass[preprintnumbers,amsmath,amssymb,aps,prd,11pt,nofootinbib]{revtex4-2}
\usepackage{color}
\newcommand*{\eps}{{\rlap{\lower2ex\hbox{$\,\,\tilde{}$}}{\epsilon_{ijk}}}}
\newcommand*{\EPS}{{\rlap{\lower2ex\hbox{$\,\,\tilde{}$}}{\epsilon_{i'j'k'}}}}
\newcommand*{\lmq}{{\rlap{\lower2ex\hbox{$\,\,\tilde{}$}}{\epsilon_{lmq}}}}
\newcommand*{\jmq}{{\rlap{\lower2ex\hbox{$\,\,\tilde{}$}}{\epsilon_{jmq}}}}
\newcommand*{\jql}{{\rlap{\lower2ex\hbox{$\,\,\tilde{}$}}{\epsilon_{jql}}}}
\newcommand*{\jlm}{{\rlap{\lower2ex\hbox{$\,\,\tilde{}$}}{\epsilon_{jlm}}}}
\newcommand*{\imq}{{\rlap{\lower2ex\hbox{$\,\,\tilde{}$}}{\epsilon_{imq}}}}
\newcommand*{\iql}{{\rlap{\lower2ex\hbox{$\,\,\tilde{}$}}{\epsilon_{iql}}}}
\newcommand*{\ilm}{{\rlap{\lower2ex\hbox{$\,\,\tilde{}$}}{\epsilon_{ilm}}}}
\newcommand*{\lmn}{{\rlap{\lower2ex\hbox{$\,\,\tilde{}$}}{\epsilon_{lmn}}}}
\newcommand*{\abc}{{\rlap{\lower2ex\hbox{$\,\,\tilde{}$}}{\epsilon_{abc}}}}
\newcommand*{\N}{{\rlap{\lower2ex\hbox{$\,\,\tilde{}$}}{N}}}
\newcommand{\tN}{{\rlap{\lower2ex\hbox{$\,\,\tilde{}$}}{N}}}
\newcommand*{\tM}{{\rlap{\lower2ex\hbox{$\,\,\tilde{}$}}{M}}}
\newcommand*{\imn}{{\rlap{\lower2ex\hbox{$\,\,\tilde{}$}}{\epsilon_{imn}}}}
\newcommand*{\qt}{\ln q^{\frac{1}{3}}}

\begin{document}
\title{New commutation relations for quantum gravity}

\author{Chopin Soo}\email{cpsoo@mail.ncku.edu.tw}
\address{Department of Physics, National Cheng Kung University, Taiwan}
\author{Hoi-Lai Yu}\email{hlyu@phys.sinica.edu.tw}
\address{Institute of Physics, Academia Sinica, Taiwan}

\bigskip

\begin{abstract}
A new set of fundamental commutation relations (CR) for quantum gravity is presented.
The basic variables are the eight components of the unimodular part of the spatial dreibein and eight $SL(3,R)$ generators which correspond to Klauder's momentric variables.
The commutation relations are not canonical, but they have well-defined group-theoretic meanings. Explicit unitary irreducible infinite-dimensional representations
are constructed both in the metric as well as dreibein representations. The dreibein components commute among themselves; but momentric, unlike momenta, do not.
This differs starkly from the usual canonical CR that allow wave functionals to be realized in either of the conjugate representations;
and it may explain why our universe seems fundamentally and intuitively ‘metric’ in nature, and not ‘conjugately realized’.
\\
\\
{Publication details: An earlier version of this work was published as\\
{\it\bf New Commutation Relations for Quantum Gravity}, \\Chopin Soo and Hoi-Lai Yu, Chin. J. Phys. {\bf 53}, 110106 (2015),\\
Chinese Journal of Physics (Volume 53, Number 6 (November 2015))
Special Issue On the occasion of 100 years since the birth of Einstein's General Relativity}
\\
\end{abstract}

\maketitle

\section{Commutation relations of geometrodynamics and momentric variables}

Successful quantization of the gravitational field remains the preeminent challenge a century after the completion of Einstein's general relativity (GR).
In quantum field theories, `equal time' fundamental commutation relations (CR) - part and parcel of causality requirements - are predicated on the existence of spacelike hypersurfaces.
Geometrodynamics bequeathed with positive-definite spatial metric is the simplest consistent framework to implement them. Expressed in terms of dreibein, $e_{ai}$, the spatial metric, $q_{ij} := \delta^{ab}e_{ai}e_{bj}$
is automatically positive-definite if the dreibein is real and non-vanishing, modulo $SO(3,C)$ gauge rotations which leave $q_{ij}$ invariant under these local Lorentz transformations. To incorporate fermions,  it is also necessary to introduce the dreibein which is more fundamental than the metric.

In usual quantum theories, the fundamental {\it canonical} CR, $[Q({\bf x}), P({\bf y})] = i\hbar\delta({\bf x},{\bf y})$, implies $\frac{P}{\hbar}$ is the generator of translations of $Q$ which are also symmetries of the `free theories' in the limit of vanishing interaction potentials.
For geometrodynamics, the corresponding canonical CR is $[q_{ij}({\bf x}), {\tilde\pi}^{kl}({\bf y})] = i\hbar\frac{1}{2}(\delta^k_i\delta^l_j + \delta^l_i\delta^k_j)\delta({\bf x},{\bf y})$. However, neither positivity of the spatial metric is preserved under arbitrary translations generated by conjugate momentum; nor is the `free theory' invariant under  translations when interactions are suppressed, because $G_{klmn}{\tilde\pi}^{kl}{\tilde\pi}^{mn}$ in the kinetic part of the Hamiltonian also contains the DeWitt supermetric\cite{DeWitt}, $G_{ijkl}$, which is dependent upon $q_{ij}$. In quantum gravity, states which are infinitely peaked at the flat metric, or for that matter any particular metric with its corresponding isometries, cannot be postulated ad hoc; consequently, the underlying symmetry of even the `free theory' is obscure.

Decomposition of the classical canonical geometrodynamic degrees of freedom, $(q_{ij},\widetilde{\pi}^{ij})$,  singles out the pair\footnote{The precise associated Poisson bracket is $\{q({\bf x}),\widetilde{\pi}({\bf y})\} = 3q\delta({\bf x},{\bf y})$.} $(\ln q^{\frac{1}{3}}$, $\widetilde{\pi})$
 which commutes with
the remaining unimodular $\overline{q}_{ij}:=q^{-\frac{1}{3}}q_{ij}$, and traceless $\overline{\pi}^{ij}:=q^{\frac{1}{3}}\bigl(\widetilde{\pi}^{ij}-\frac{1}{3}q^{ij}\widetilde{\pi}\bigr)$ variables.
Hodge decomposition for compact manifolds yields
$\delta \qt = \delta{T}+\nabla_i\delta{Y}^i$, wherein the spatially-independent $\delta T$ is a three-dimensional diffeomorphism invariant (3dDI) entity which serves as the intrinsic time interval, whereas $\nabla_i\delta{Y}^i$ can be gauged away since
${\mathcal L}_{\delta{{\overrightarrow N}}}\qt = \frac{2}{3}\nabla_i{\delta N^i}$. A theory of quantum geometrodynamics dictated by first-order Schr\"{o}dinger evolution
 in intrinsic time, $i\hbar\frac{d\Psi}{d{T}}={H}_{\rm Phys.}\Psi$,  and
equipped with diffeomorphism-invariant physical Hamiltonian and time-ordering has been discussed and advocated in Refs.\cite{SOOYU, SOOYU1, SOO3,ITG,ITGBook} within this Intrinsic Time Geometrodynamics framework.
The Hamiltonian, ${H}_{\rm Phys}=\int\frac{{\bar H}({\bf x})}{\beta}d^3{\bf x}$, and ordering of the time development operator,
$U(T,T_0)={\bf T}\{{\exp}[-\frac{i}{\hbar}\int^T_{T_0}H_{\rm Phys}(T')\delta{T}']\}$, are 3dDI provided
\begin{eqnarray}
\label{INTRINSIC1}
{\bar H} = {\sqrt{ {\bar \pi^{ij}}  {\bar G_{ijkl}} {\bar \pi^{kl}} + \mathcal{V}[q_{ij}]}}
\end{eqnarray}\noindent
is a scalar density of weight one; and Einstein's GR (with $\beta= \frac{1}{\sqrt{6}}$ and ${\mathcal V} =- \frac{q}{(2\kappa)^2}[R - 2\Lambda_{\it{eff}} $]) is a particular realization of this wider class of theories. Natural extensions of GR and the precise quantum Hamiltonian are discussed further in Refs.\cite{SOOYU, SOO3,ITGBook}.

The Poisson brackets for the barred variables are,
\begin{eqnarray}\label{qT}
\{{\bar{q}_{ij}({\bf x}),\bar{q}_{kl}({\bf y})\} =0,\,
\{\bar{q}_{kl}({\bf x}),{\bar{\pi}}^{ij}({\bf y})\}= P^{ij}_{kl}\,\delta({\bf x},{\bf y}),\, \{\bar{\pi}}^{ij}({\bf x}),{\bar{\pi}}^{kl}({\bf y})\}=\frac{1}{3}({\bar q}^{kl}{\bar{\pi}}^{ij} -{\bar q}^{ij}{\bar{\pi}}^{kl})\delta({\bf x},{\bf y});
\end{eqnarray}
with $P^{ij}_{kl} :=  \frac{1}{2}(\delta^i_k\delta^j_l + \delta^i_l\delta^j_k) - \frac{1}{3}\bar{q}^{ij}\bar{q}_{kl}$ denoting the traceless projection operator. This set is not strictly canonical.
In the metric representation, the implementation of ${\bar{\pi}}^{kl}$ as traceless, symmetric, and self-adjoint operators is problematic. Remarkably, these difficulties can be cured by passing to the `momentric variable' (first introduced by Klauder\cite{Klauder}) which is classically $\bar \pi^{i}_{j} = \bar q_{jm}\bar \pi^{im}$.
In terms of spatial metric and momentric variables, the fundamental CR postulated (from which the classical Poisson brackets corresponding to \eqref{qT} can be recovered)  are then\cite{SOO3}
\begin{eqnarray}
&&[ \bar q_{ij}({\bf x}), \bar q_{kl}({\bf y})]=0,\quad
[\bar q_{ij}({\bf x}), {\bar{\pi}}^{k}_{l}({\bf y})]= i\hbar\bar{E}^k_{l(ij)}\delta({\bf x},{\bf y}),\cr
&&[ {\bar{\pi}}^{i}_{j}({\bf x}), {\bar{\pi}}^{k}_{l}({\bf y})]= \frac{i\hbar}{2}\bigl(\delta^k_j{\bar{\pi}}^i_l-\delta^i_l{\bar{\pi}}^k_j\bigr)\delta({\bf x},{\bf y});\label{REL}
\end{eqnarray}\noindent
wherein $\bar{E}^i_{j(mn)}:=\frac{1}{2}\bigl(\delta^i_m\overline{q}_{jn}+\delta^i_n\overline{q}_{jm}\bigr)-\frac{1}{3}\delta^i_j\overline{q}_{mn}$
(with properties $\delta^{j}_{i}\bar{E}^i_{j(mn)} = \bar{E}^i_{j(mn)} \bar q^{mn}=0;\bar{E}^i_{j(il)}=\bar{E}^i_{j(li)}=\frac{5}{3}\overline{q}_{jl}$) is the vielbein for the
supermetric ${\bar G}_{ijkl} =  \bar{E}^m_{n(ij)}\bar{E}^n_{m(kl)}$.
Quantum mechanically, the momentric operators and CR {\it can be explicitly realized  in the metric representation} by
\begin{equation}\label{pp}
{\bar{ \pi}}^{i}_{j}({\bf x}):=\frac{\hbar}{i}\bar{E}^i_{j(mn)}({\bf x})\frac{\delta}{\delta \bar q_{mn}({\bf x})} =\frac{\hbar}{i}\frac{\delta}{\delta \bar q_{mn}({\bf x})}\bar{E}^i_{j(mn)}({\bf x})=\hat{\bar{ \pi}}^{\dagger i}_{j}({\bf x})
\end{equation}\noindent
which are self-adjoint on account of $[\frac{\delta}{\delta\bar{q}_{mn}({\bf x})},\bar{E}^i_{j(mn)}({\bf x})]=0$.

\section{New commutation relations}
These momentric variables generate $SL(3,R)$ transformations of $\bar{q}_{ij} =\delta^{ab}{\bar e}_{ai}{\bar e}_{bj} $ and dreibein ${\bar e}_{ai}$ which preserve the positivity and unimodularity of this spatial metric. Moreover, the commutation relations \eqref{REL} for the self-adjoint momentric indicate they generate a unitary infinite-dimensional algebra\cite{Rosen} that is, at each spatial point, a (non-compact) $sl(3,R)$ algebra. $sl(3,R)$ and $su(3)$ share the same complexification (to $sl(3,C)$) even though they are not isomorphic as real Lie algebras. In fact, with $3\times 3$ Gell-Mann matrices $\lambda^{A=1,...,8}$,  it can be checked that $T^{A}({\bf x}):= \frac{1}{\hbar}(\lambda^{A})^{j}_{i}\hat{\bar \pi}^{i}_{j}({\bf x})$  generate an algebra with the same structure constants $f^{AB}\,_C$ of $su(3)$\cite{Gell-Mann}.

In \eqref{REL} there is an asymmetry in that there are only five independent components in the symmetric  unimodular ${\bar q}_{ij}$ (likewise for the symmetric traceless $\bar\pi^{ij}$  in \eqref{qT}),  whereas the mixed-index traceless momentric variable, ${\bar \pi}^i_j$, contains eight components. This asymmetry is rectified by  unimodular dreibein-traceless momentric variables, $({\bar e}_{ai} := e^{-\frac{1}{3}}e_{ai}, {\bar{ \pi}}^{k}_{l})$, each having eight independent components. To incorporate Weyl fermions and gauge the local $SL(2,C)$ Lorentz group, it is necessary to introduce the dreibein\cite{ITGBook}. The dreibein-momentric variables obey the {\it new fundamental CR\cite{ITGBook} advocated in this work},\footnote{An equivalent set of CR is \begin{equation}
[{\bar e}_{ai}({\bf x}), {\bar e}_{bj}({\bf y})] =0,\quad [{\bar e}_{ai}({\bf x}), T^A({\bf y})] =i(\frac{\lambda^A}{2})^k_i{\bar e}_{ak}{\delta({\bf x},{\bf y})},\quad
[T^{A}({\bf x}),T^{B}({\bf y})]= -{f}^{AB}\,_CT^{C}{\delta({\bf x},{\bf y})\nonumber}.
\end{equation}
These CR have clear group-theoretic meanings, and neither the gravitational coupling constant nor Planck's constant make their appearance. }
\begin{eqnarray}\label{fundamental}
&[{\bar e}_{ai}({\bf x}), {\bar e}_{bj}({\bf y})] =0,\quad [{\bar e}_{ai}({\bf x}),{\bar{\pi}}^{j}_{k}({\bf y})] =\frac{i\hbar}{2}(\delta^j_i{\bar e}_{ak}-\frac{1}{3}\delta^j_k{\bar e}_{ai}){\delta({\bf x},{\bf y})},\cr
& [ {\bar{\pi}}^{i}_{j}({\bf x}), {\bar{\pi}}^{k}_{l}({\bf y})]= \frac{i\hbar}{2}(\delta^k_j{\bar{\pi}}^i_l-\delta^i_l{\bar{\pi}}^k_j)\delta({\bf x},{\bf y}).
\end{eqnarray}
The second CR in \eqref{fundamental} implies
\begin{equation}
U^\dagger(\alpha){\bar e}_{ai}({\bf x})U(\alpha) = \Big[\exp({\frac{\alpha({\bf x})}{2}})\Big]^j_i{\bar e}_{ja}({\bf x}),
\end{equation}
with $ U(\alpha) :=\exp({-\frac{i}{\hbar}\int \alpha^i_j({\bf y}){\bar \pi}^j_i ({\bf y})d^3{\bf y}})$ and traceless $\alpha^i_j$. This is just a local $SL(3,R)$ transformation of the dreibein via matrix multiplication with $\exp({\frac{\alpha({\bf x})}{2}})$; while the final CR in \eqref{fundamental} is the statement that the momentric ${\bar \pi}^j_i$  generates, at each spatial point, a separate $sl(3,R)$ algebra.

Local Lorentz transformations are generated by the Gauss Law constraint $G_a({\bf x}):= \epsilon_{abc}e^b_i{\tilde \pi}^{ci}-i{\tilde\pi}_\psi\frac{\tau_a}{2}\psi=0$, wherein $\tilde\pi^{ai}$ (which is classically $\tilde\pi^{ai}=2{\tilde\pi}^{ij}e^a_{j}$) is the canonical conjugate momentum of the dreibein, $ (\psi, {\tilde\pi}_\psi)$ denotes the conjugate pair for a Weyl fermion, and $\tau_a$ are the Pauli matrices. However, $e:=\det(e_{ai})$ and $(\ln q^{\frac{1}{3}}$, $\widetilde{\pi})$ are Lorentz singlets which commute with the Gauss Law constraint; and the constraint can be reexpressed with traceless momentric, rather than momentum, variable as $G_a = \epsilon_{abc}{\bar e}^b_{i}{\bar e}^{cj}\bar\pi^i_j -i{\tilde\pi}_\psi\frac{\tau_a}{2}\psi =0$. That it generates $SO(3,C)$ local Lorentz rotations of the dreibein can be verified from $[ {\bar e}_{ai}({\bf x}), \frac{i}{\hbar}\int \eta^b G_b d^3{\bf y} ] = \epsilon_{abc}\eta^b({\bf x}){\bar e}^c_{i}({\bf x})$, with $\eta^b$ as gauge parameter; in addition, it generates the associated local $SL(2,C)$ Lorentz transformations of the fermionic degrees of freedom (d.o.f.).

The momentric generators, unlike the usual momentum variables, obey \eqref{fundamental}, and thus do not commute among themselves. A remarkable consequence is that quantum wave functions cannot be chosen as simultaneous functionals of all the ${\bar \pi}^j_i$; {\it in contradistinction, quantization in the complete metric or dreibein representation is viable, and in this context `preferentially selected' by the fundamental CR}.  This differs starkly from the usual  {\it canonical } CR which allows wave functionals to be realized in either of the conjugate representations; and it may explain why our universe seems  fundamentally and intuitively `metric' in nature, and not `conjugately realized'. The dreibein (and metric representation) consistently guarantees that hypersurfaces on which the fundamental CR are defined will be spacelike\cite{ITGBook}.

An explicit realization of the momentric operator in the dreibein representation is
\begin{equation}
{\bar\pi}^i_j = \frac{\hbar}{2i}({\bar e}_{aj}\frac{\delta}{\delta{\bar e}_{ai}} -\frac{1}{3}\delta^i_j{\bar e}_{ak}\frac{\delta}{\delta{\bar e}_{ak}}).
\end{equation}
This is self-adjoint and realizes \eqref{fundamental} when operating on wave functionals of the dreibein.  It is a unitary infinite-dimensional irreducible representation of non-compact SL(3,R); and isomorphic to the representation presented in Ref.\cite{Rosen} which also rigorously demonstrated the action of these generators on well-defined Hilbert spaces.

The symmetry of the free theory now becomes transparent. It is characterized by $SL(3,R)$ invariance generated by the momentric, because the kinetic operator in \eqref{INTRINSIC1} can now be interpreted to be the $SL(3,R)$ Casimir,
\begin{equation}
{{\bar \pi}^{ij}}  {{\bar G}_{ijkl}} { {\bar \pi}}^{kl}={{\bar \pi}}^{i}_{j}{{\bar \pi}}^{j}_{i} ={{\bar \pi}}^{i\dagger }_{j} {{\bar \pi}}^{j}_{i}.
\end{equation}
\noindent
The spectrum of the free Hamiltonian and quantum states can thus be labeled by eigenvalues $\{h, d, j, m\}$ corresponding to the commuting set at each spatial point comprising the two $SL(3,R)$
Casimirs (${{\bar \pi}}^{i}_{j}{{\bar \pi}}^{j}_{i}$ and $\det({\bar \pi}^{i}_{j})$),  the Casimir $J^2$ of the embedded $SO(3)$ and its corresponding $J_3$.
An underlying group structure has the advantage the action of momentric on wave functionals  by functional differentiation can be traded for the well-defined action of generators on quantum states expanded in this basis. To wit,
\begin{equation}
\label{CASIMIR}
\frac{\hbar}{2i}({\bar e}_{aj}({\bf x})\frac{\delta}{\delta{\bar e}_{ai}({\bf x})} -\frac{1}{3}\delta^i_j{\bar e}_{ak}({\bf x})\frac{\delta}{\delta{\bar e}_{ak}({\bf x})})
\langle\bar{e}_{bl}|\prod_{\bf y}{|h, d, j, m\rangle}_{\bf y}
=\langle\bar{e}_{bl}|{\bar\pi}^i_j ({\bf x})\prod_{\bf y}{|h, d, j, m\rangle}_{\bf y} .
\end{equation}


 Group theoretical considerations also lead to a succinct description of graviton d.o.f. and their associated quantum excitations.
  In geometrodynamics, local $SL(3,R)$ transformations of ${\bar q}_{kl}$ are generated through\cite{Klauder} $U^\dagger(\alpha) {\bar q}_{kl}({\bf x}) U(\alpha) = (e^{\frac{\alpha({\bf x})}{2}})^m_k {\bar q}_{mn}({\bf x}) (e^{\frac{\alpha({\bf x})}{2}})^n_l$; while the generator of spatial diffeomorphisms for the momentric and unimodular spatial d.o.f. is effectively $ D_i =-2{\nabla}_j{\bar\pi}^j_i$; consequently,
with smearing, $\int N^i D_i d^3{\bf x} =\int (2{\nabla}_jN^i){\bar\pi}^j_i  d^3{\bf x} $ after integration by parts. The action of spatial diffeomorphisms can thus be subsumed by specialization to $\alpha^i_j =2\nabla_jN^i$, with the upshot
that $SL(3,R)$ transformations which are {\it not} spatial diffeomorphisms are parametrized by $\alpha^i_j$ complement to $2{\nabla}_jN^i$. Given a background metric ${q}^{\cal B}_{ij}= q^{\frac{1}{3}}{\bar q}^{\cal B}_{ij}$, this complement is
 precisely characterized by the choice of  transverse traceless (TT) parameter  $(\alpha_{TT})^i_j  := {q}^{\cal B}_{jk}\alpha^{(ik)}_{Phys}$, because the condition $\nabla_{\cal B}^j(\alpha_{TT})^i_j=0 $ excludes non-trivial $N^i$  through $\nabla^2_{\cal B}N^i=0$ if $(\alpha_{TT})^i_j $ were of the form $2\nabla^{\cal B}_j N^i$. The TT conditions impose four restrictions on the symmetric $\alpha^{(ij)}_{Phys}({\bf x})$, leaving exactly two free parameters. The action of $U_{Phys}( \alpha_{TT}) = e^{-\frac{i}{\hbar}\int (\alpha_{TT})^i_j {\bar \pi}^j_i d^3{\bf x}}$  (which is thus a local $SL(3,R)$ modulo spatial diffeomorphism) on any 3dDI wave functional would result in an inequivalent state. TT conditions however require a particular background metric to be defined. In Ref.\cite{SOOITA} a basis of infinitely squeezed states was explicitly realized by Gaussian wave functionals $\Psi[\bar q]_{{q}^{\cal B}} \propto \exp[-\frac{1}{2}\int {\tilde f}_{\epsilon}({\bar q}_{ij}-{\bar q}^{\cal B}_{ij}) {\bar G}^{ijkl}_{\cal B} ({\bar q}_{kl}-{\bar q}^{\cal B}_{kl}) d^3{\bf x}]$.  3dDI is recovered in the limit of zero Gaussian width with divergent $\hbox{lim}_{\epsilon\rightarrow{0}}{\tilde f}_{\epsilon}\rightarrow \hbox{lim}_{{\bf x}\rightarrow {\bf x'}}\delta({\bf x'},{\bf x})$. These  localized Newton-Wigner states are infinitely peaked at $q^B_{ij}$ which can then be deployed to actualize the TT conditions.  The action of $U_{Phys}( \alpha_{TT})$ on these states would thus generate two infinitesimal local physical excitations at each spatial point.

\section*{Acknowledgments}

This work was supported in part by the Ministry of Science and Technology (R.O.C.) under Grant Nos.
NSC101-2112-M-006 -007-MY3 and MOST104-2112-M-006-003, and the Institute of Physics, Academia Sinica.


\begin{thebibliography}{99}


\bibitem{DeWitt}Bryce S. DeWitt, Phys. Rev. {\bf 160}, 1113 (1967).

\bibitem{SOOYU}{C. Soo and H.-L. Yu,
{\it General relativity without the paradigm of space-time covariance and resolution of the problem of time},
Prog. Theor. Exp. Phys.,  013E01 (2014)}.

\bibitem{SOOYU1} {N. \'{O} Murchadha, C. Soo and H.-L. Yu,
{\it Intrinsic time gravity and the Lichnerowicz-York equation},
Class. Quantum Grav. {\bf  30},  095016 (2013)}.

\bibitem{SOO3}
E. E. Ita III, C. Soo, H.-L. Yu, {\it Intrinsic time quantum geometrodynamics}, Prog. Theor. Exp. Phys., 083E01 (2015).

\bibitem{ITG}Huei-Chen Lin and Chopin Soo, {\it Intrinsic time geometrodynamics: explicit examples}, Chin. J. Phys.  53, 110102 (2015)
(Chinese Journal of Physics Special Issue On the occasion of 100 years since the birth of Einstein's General Relativity).

\bibitem{ITGBook}Chopin Soo and Hoi-Lai Yu, {\it Intrinsic Time Geometrodynamics: At One With The Universe}
(World Scientific Publishing Co., 2022)  (https://doi.org/10.1142/13062).

\bibitem{Klauder} J. R. Klauder, {\it Overview of affine quantum gravity}, Int. J. Geom. Meth. Mod. Phys. {\bf 3}, 81 (2006), and references therein.

\bibitem{Rosen}G. Rosen, {\it Unitary Irreducible Representations of $SL(3, R)$},  J.  Math. Phys. {\bf 7}, 1284 (1966).

\bibitem{Gell-Mann}{\it The eightfold way}, M. Gell-Mann and Y. Ne'eman (W. A. Benjamin, 1964).

\bibitem{SOOITA}E. Ita and C. Soo,
Annals of Physics {\bf 309}, 80  (2015).







\end{thebibliography}
\end{document}